\documentclass[prl,superscriptaddress,preprint,endfloats]{revtex4}
\usepackage[dvipdfmx]{graphicx}
\usepackage{bm} 
\usepackage{pifont}
\usepackage{amsmath, amsthm, amssymb}
\usepackage{color}
\usepackage{dcolumn} 
\begin{document}
\title{Near-half-metallic state in the half Heusler PtMnSb film on a III-V substrate}
\author{Shinichi Nishihaya}
\email[]{nishihaya@phys.sci.isct.ac.jp}
\altaffiliation{Present Address: Department of Physics, Institute of Science Tokyo, Tokyo, 152-8551, Japan} 
\affiliation{Department of Electrical and Computer Engineering, University of California Santa Barbara, Santa Barbara, California 93106, USA}
\author{Malcolm J. A. Jardine}
\affiliation{Department of Materials Science and Engineering, Carnegie Mellon University, Pittsburgh, Pennsylvania 15213, USA}
\author{Hadass S. Inbar}
\affiliation{Materials Department, University of California Santa Barbara, Santa Barbara, California 93106, USA}
\author{Aranya Goswami}
\altaffiliation{Present Address: Department of Electrical Engineering and Computer Science, Massachusetts Institute of Technology, Cambridge, MA 02139, USA}
\affiliation{Department of Electrical and Computer Engineering, University of California Santa Barbara, Santa Barbara, California 93106, USA}
\author{Jason T. Dong}
\affiliation{Materials Department, University of California Santa Barbara, Santa Barbara, California 93106, USA}
\author{Aaron N. Engel}
\affiliation{Materials Department, University of California Santa Barbara, Santa Barbara, California 93106, USA}
\author{Yu-Hao Chang}
\affiliation{Materials Department, University of California Santa Barbara, Santa Barbara, California 93106, USA}
\author{Connor P. Dempsey}
\affiliation{Materials Department, University of California Santa Barbara, Santa Barbara, California 93106, USA}
\author{Makoto Hashimoto}
\affiliation{Stanford Synchrotron Radiation Lightsource, SLAC National Accelerator Laboratory,
2575 Sand Hill Road, Menlo Park, California 94025, USA}
\author{Donghui Lu}
\affiliation{Stanford Synchrotron Radiation Lightsource, SLAC National Accelerator Laboratory,
2575 Sand Hill Road, Menlo Park, California 94025, USA}
\author{Noa Marom}
\affiliation{Department of Materials Science and Engineering, Carnegie Mellon University, Pittsburgh, Pennsylvania 15213, USA}
\affiliation{Department of Physics and Department of Chemistry, Carnegie Mellon University, Pittsburgh, Pennsylvania 15213, USA}
\author{Chris J. Palmstr\o m}
\email[]{cjpalm@ucsb.edu}
\affiliation{Department of Electrical and Computer Engineering, University of California Santa Barbara, Santa Barbara, California 93106, USA}
\affiliation{Materials Department, University of California Santa Barbara, Santa Barbara, California 93106, USA}

\begin{abstract}
The interplay between half-metallic ferromagnetism and spin-orbit coupling within the inversion-symmetry-broken structure of half Heuslers provides an ideal platform for various spintronics functionalities. Taking advantage of good lattice matching, it is highly desired to epitaxially integrate promising Heuslers into III-V semiconductor-based devices. PtMnSb is one of the first half Heuslers predicted to be an above-room-temperature half-metal with large spin orbit coupling, however, its half-metallicity and potential as a spintronics material has remained elusive due to lack of high quality samples. Here we demonstrate epitaxial growth of single crystal PtMnSb(001) film on GaSb(001) substrates using molecular beam epitaxy. \color{black}Direct observation of the band structure via angle-resolved photoemission spectroscopy and many-body perturbation theory within the quasi-particle self-consistent GW approximation (QPGW) reveal that PtMnSb hosts rather a near-half-metallic state with both spin bands crossing the Fermi level and with high spin polarization over 90\%. \color{black}Temperature dependence of magnetization also shows an anomalous enhancement below 60 K, which can be associated with the development of such a near-half-metallic state at low temperatures. \color{black}Epitaxial growth of high crystalline PtMnSb on a III-V paves the way for systematic clarification of its spin transport properties with fine-tuning of strain in heterostructure devices.
\end{abstract}
\maketitle
\section{I. INTRODUCTION}
Half-Heuslers represented by $XYZ$ ($X$ is transition or rare earth metals, $Y$ transition metals, and $Z$ main group metals) form a large family of ternary intermetallic compounds and realize rich electronic and magnetic phases depending on the vast combination of constituent elements in the same crystal structure \cite{Heusler_rev1,Heusler_rev2,Heusler_rev3}. Their tunable properties via chemical substitution as well as good lattice-matching to the major III-V semiconductor compounds have made Heuslers as an ideal platform for integrating multiple functionalities to existing semiconductor-based electronics \cite{Heusler_rev2}. Among others, a particular research interest has been devoted to exploration of above-room-temperature half-metallicity, which has been mostly found in the Heuslers based on 3$d$ transition metals. The unique electronic structure exhibiting an ideally 100\% polarized spin state at the Fermi level has found useful application opportunities in spintronics, such as efficient spin injectors, spin valves, and magnetic tunnel junctions \cite{Heusler_rev3}. In addition to the half-metallicity, the inversion-symmetry-broken structure of half-Heusler $XYZ$ compounds also opens a new avenue for spin-dependent transport phenomena enabled by the relativistic spin-orbit coupling (SOC), leading to promising functionalities including spin-orbit torques, unidirectional magnetoreisistance, and spin-charge conversion as recently demonstrated for the half-metallic half-Heusler NiMnSb \cite{NMS_exp1,NMS_exp2,NMS_exp3}. 

Among many half-metallic half-Heuslers found to date \cite{Heusler_rev1,Heusler_rev2,Heusler_rev3}, PtMnSb, together with NiMnSb, was the first-predicted half-metal ferromagnet with a high Curie temperature ($T_{\mathrm{C}}$ = 582 K) \cite{PMS_cal,PMS_Tc}. Due to inclusion of heavy metal Pt, SOC is expected to play an important role in the half-metallicity of PtMnSb \cite{PMS_SOC_cal}, which is in contrast to NiMnSb. The energy bands near the Fermi level are dominated by Pt 6$p$ orbital character and show a large SOC splitting. The half-metal gap has been predicted to depend sensitively on the lattice parameter \cite{PMS_SOC_cal,PMS_strain1}. Experimentally, PtMnSb has gathered attention rather in the context of magneto-optical applications due to its exceptionally large magneto-optical Kerr rotation angle at room temperature \cite{PMS_MO_cal,PMS_MO_exp,PMS_MO_exp2,PMS_MO_exp3,PMS_MO_exp4}. Despite the promising properties expected for PtMnSb with large SOC, \color{black}however, its half-metallicity and potential as a spintronics material have been elusive. \color{black}This is mainly attributed to lack of high-quality PtMnSb films and heterostructures. While film growth of PtMnSb has been endeavored so far on oxide substrates and on metal buffers \cite{PMS_film1,PMS_film2,PMS_film3}, stabilizing single crystal PtMnSb has been challenging. This has limited previous studies to either polycrsytalline films or low-crystalline quality ones grown at low temperatures. Considering the epitaxial and chemical compatibility of the half-Heusler structure with III-Vs, as well as the flexible lattice constant tuning available on the major III-V semiconductor substrates, it is highly desirable to achieve epitaxial PtMnSb films on III-Vs for further investigation of its spintronics functionalities. 

\begin{figure*}
\begin{center}
\includegraphics[width=14cm]{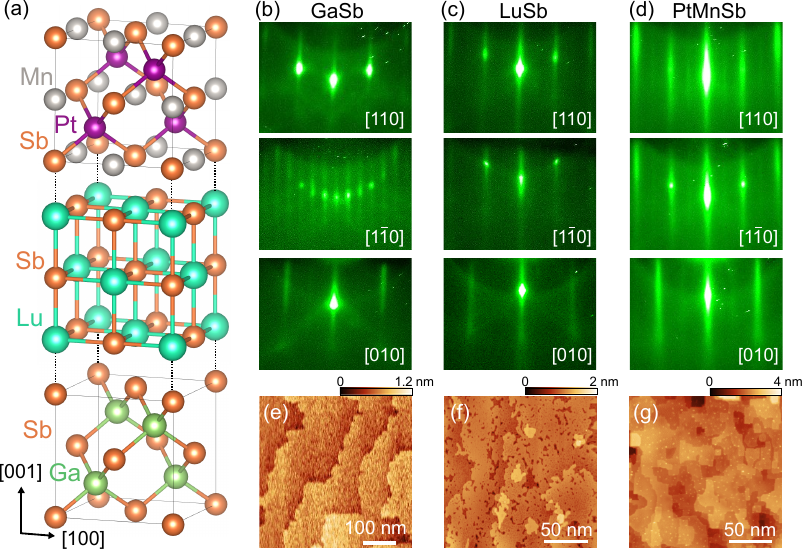}
\caption{Surface reconstruction and morphology at each growth step. (a) Schematic of crystal structure of PtMnSb/LuSb/GaSb stacked along [001]. RHEED patterns along [110], [1$\bar{1}$0], and [010] azimuth for (b) GaSb, (c) LuSb, and (d) PtMnSb surfaces. (e)-(g) Surface morphology images taken by \textit{in-situ} STM for each layer. 
}
\label{fig1}
\end{center}
\end{figure*}

In this study, we report for the first time to our knowledge, epitaxial stabilization of the half Heusler PtMnSb on a III-V subtrate by molecular beam epitaxy. By adopting a thin layer of rock-salt type LuSb as a barrier for preventing the interfacial diffusion, PtMnSb (001) films of high crystalline quality are successfully grown on GaSb(001). \color{black}Via angle-resolved photoemission spectroscopy (ARPES) measurements and many-body perturbation theory calculations employing quasi-particle self-consistent GW (QPGW), we have clarified that PtMnSb is not a pure half-metal but rather a near-half-metal with high spin polarization at the Fermi level. Magnetization and transport properties of PtMnSb films were evaluated down to cryogenic temperatures, where signatures consistent with such a near-half-metal state have been observed.\color{black}

\section{II. METHODS}
\subsection{MBE growth}
GaSb(001) ($a = 6.096$\AA) wafers with a relatively small lattice mismatch to PtMnSb ($a = 6.21$\AA) were chosen as substrates. Prior to the growth of LuSb and PtMnSb, a 500 nm thick GaSb buffer layer was first grown on GaSb(001) substrates in a modified VG V80H III-V MBE chamber (base pressure $< 5 \times 10^{-11}$ Torr). The native oxide layer on the GaSb substrates was thermally desorbed under Sb$_2$ overpressure at 550$^{\circ}$C measured by an infrared pyrometer (emissivity set to 0.62-0.67), which is also used to calibrate the temperature for the GaSb buffer growth under Sb$_2$ overpressure at 530$^{\circ}$C. To avoid the quality variation of the substrate preparation, we have performed this GaSb buffer growth on a 2 inch diameter GaSb wafer, capped it with approximately 1 $\mu$m thick Sb, and cleaved it into smaller pieces \textit{ex-situ} for different growths \cite{Sbdecapping}. Each of the cleaved samples were then transferred to the Veeco MOD Gen II chamber, and decapped by annealing at 450$^{\circ}$C under Sb$_4$ overpressure for subsequent LuSb and PtMnSb growth. The surface of the GaSb(001) buffer after Sb decapping exhibits a Sb-rich c(2$\times$6) reconstruction (Fig. 1(b)) and a flat surface with clear step and terrace structures (Fig. 1(e)). 

LuSb barrier layers were grown at 300$^{\circ}$C by co-deposition of Lu and Sb using an effusion cell with a slightly Sb-rich flux ratio (Lu:Sb = 1:1.1) which was calibrated in advance by Rutherford backscattering spectrometry (RBS) measurements of calibration structures grown on Si substrates. LuSb(001) surface shows an unreconstructed (1$\times$1) pattern (Fig. 1(c)). The growth rate of LuSb was 1.2 \AA /min deduced from the RHEED intensity oscillation period, and the typical thickness was 8 monolayers (ML). In order to evaporate the excess Sb and realize a flat surface of the LuSb layer (Fig. 1(f)), post-growth annealing was performed at 330-350$^{\circ}$C for 30 min.

PtMnSb films were grown by first nucleating a 4-unit-cell-thick seed layer at 225$^{\circ}$C by alternately supplying Pt and Mn+Sb flux in a timed shutter sequence starting from Pt deposition. Pt was evaporated from an electron beam evaporator in an effusion cell port, while Mn and Sb were supplied from effusion cells. The shutter open period for depositing each Pt and MnSb layer was calculated based on the RBS-calibrated atomic flux of each element. The remainder of PtMnSb film were grown at 300$^{\circ}$C by co-deposition of Pt, Mn and Sb with a 1:1:1.1 ratio, followed by annealing at 350$^{\circ}$C under Sb$_4$ overpressure. Before sample removal for \textit{ex-situ} characterization, PtMnSb films were capped with 3-10 nm thick AlO$_x$ by e-beam evaporation of Al$_2$O$_3$ to avoid oxidization. The PtMnSb film used for ARPES measurements were grown on Zn-doped GaSb(001) substrate with a 500 nm thick Be-doped GaSb buffer layer following the same growth procedures.

\begin{figure*}
\begin{center}
\includegraphics[width=16cm]{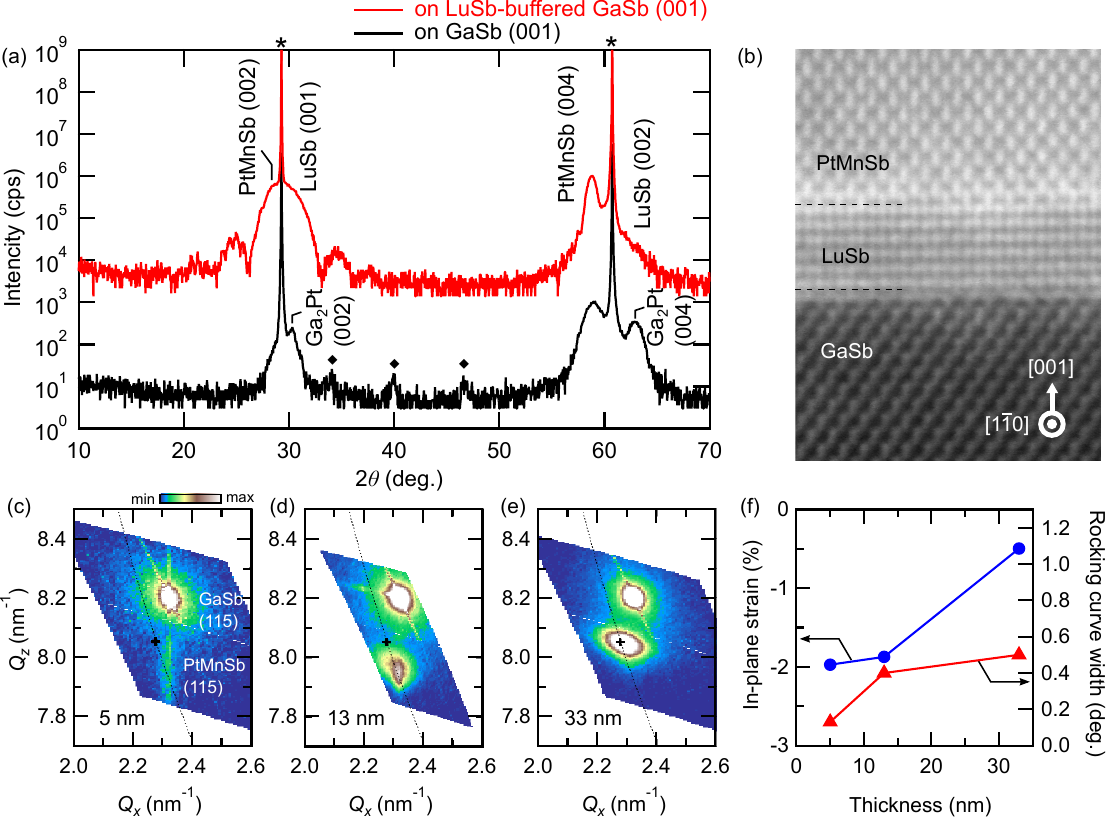}
\caption{
Structural characterization of the PtMnSb films. (a) XRD $\theta$-2$\theta$ scans for PtMnSb films grown directly on GaSb(001) and LuSb-buffered GaSb(001). The asterisk marks the diffraction peaks from GaSb substrate, and the diamond the impurity peaks formed as result of decomposition of PtMnSb. (b) Cross-sectional STEM image along the [1$\bar{1}$0] direction of the PtMnSb/LuSb/GaSb structure. Reciprocal space maps collected around the PtMnSb (115) Bragg peak for (c) 5 nm, (d) 13 nm, and (e) 33 nm thick PtMnSb films. The cross mark shows the diffraction peak position for bulk PtMnSb, and the dotted line indicates the lattice parameter changes following a Poisson ratio of 0.26. (f) Thickness dependence of in-plane compressive strain (left axis) and rocking curve width of PtMnSb (004) peak (right axis).
}
\label{fig2}
\end{center}
\end{figure*}

\subsection{Characterization}
\color{black}The surface morphology at each growth step were characterized by reflection high-energy elctron diffraction (RHEED), and by in-situ scanning tunneling microscopy (STM) performed in an Omicron VT-STM at room temperature with
a bias of 2 V and a tunneling current of 100 pA. Structural properties were characterized by x-ray diffraction (XRD) and scanning transmission electron microscopy (STEM). Magnetization measurements at temperatures were performed on a SQUID magnetometer (MPMS, Quantum Design). Temperature dependence of magnetization were measured between 2 to 300 K with applied magnetic field of 0.1 T following the initial zero-field cooling. Transport measurements with magnetic field up to 14 T were performed in a 4He cryostat
(PPMS, Quantum Design). A standard low-frequency ac lock-in technique was used with a constant excitation current of 10 $\mu$A. 

\color{black}ARPES measurements were performed under 20 K at beamline 5-2 of the Stanford Synchrotron Radiation Lightsource (SSRL) using linear-polarized light. Spectra were collected with a ScientaOmicron DA30L electron analyzer. PtMnSb films were transferred in vacuo from the MBE system at UCSB to SSRL via a vacuum suitcase with base pressure
better than $4\times 10^{-11}$ Torr. \color{black}

\subsection{Computational details}
\color{black}For the purpose of comparison with the ARPES measurements conducted here, the $a$ and $b$ lattice parameters of PtMnSb were set to 6.096 \AA, to match the GaSb substrate. This corresponding to a biaxial compressive strain of 1.9 \%. Accordingly, the $c$ lattice parameter was increased to 6.290 \AA\ using a Poisson ratio of 0.26 (See Supplemental Material for the strain dependence of the electronic structure \cite{SM}).
All calculations were performed using the Vienna Ab initio Simulation Package (VASP) with the projector augmented wave (PAW) method \cite{Cal_1,Cal_2,Cal_3}. A semi-local DFT calculation was performed using the generalized gradient approximation (GGA) of Perdew, Burke, and Ernzerhof (PBE) \cite{Cal_4, Cal_5}. The QPGW \cite{Cal_6,Cal_7,Cal_sup} calculations were performed with the range-separated hybrid functional of Heyd-Scuseria-Ernzerhof (HSE) \cite{Cal_8,Cal_9} used as the starting point. 
All calculations used a plane-wave basis set with a kinetic energy cutoff of 300 eV and an energy convergence criterion of $10^{-6}$ eV. Spin-orbit coupling (SOC) was used throughout (an analysis of the effect of SOC on the electronic structure is provided in the Supplemental Material \cite{SM}). Convergence parameters were set to BMIX = 3, AMIN = 0.01, and ALGO = Fast.
The recommended pseudopotentials provided by VASP were used for PBE calculations and the Pt\_GW, Mn\_GW and  Sb\_sv\_GW pseudopotentials were used for QPGW calculations.
Based on convergence tests provided in the Supplemental Material \cite{SM}, four self-consistent steps were performed for QPGW calculations. 72 available bands in the virtual orbitals provided a sufficient number of unoccupied orbitals. The number of bands for which the self-energy shift was calculated (NBANDSGW) was set to 30. The number of imaginary frequency and imaginary time grid points, NOMEGA, was set to 100. Regarding the convergence of QPGW calculations, we note that the settings used here are adequate close to the Fermi level, which is the energy range of interest to us, however tighter convergence settings, in particular for NOMEGA and NBANDSGW, may be needed if states farther than about $\pm$2 eV from the Fermi level are of interest. QPGW band structures were calculated via Wannier interpolation using the Wannier90 code \cite{Cal_10}.  The Brillouin zone was sampled using the $\Gamma$-centered scheme. For QPGW and HSE calculations a $5 \times 5 \times 5$ $k$-point mesh was used. For PBE calculations a k-point mesh of $7\times 7\times 7$ was used for the self-consistent field (SCF) cycle and a k-point mesh of $11\times 11\times 11$ was used for the density of states (DOS). The magnetic moment was read from the magnetization output in the OUTCAR of a self consistent calculation with LORBIT=11. The Spin Polarization percentage (SP\%) at the Fermi level was calculated as follows:
\begin{equation}
{P (E_{f})}=\frac{D_{\uparrow}(E_{f}) - D_{\downarrow}(E_{f})}{D_{\uparrow}(E_{f}) + D_{\downarrow}(E_{f})}
\end{equation}
where D$_{\uparrow}$ and D$_{\downarrow}$ are the spin-up and spin-down density of states in a 0.075 eV window around the Fermi level. 
The Fermi surface was calculated using the open-source Python package, iFermi \cite{Cal_11}. Comparisons of QPGW, with and without bi-axial strain, are given in the Supplemental Material \cite{SM}.
Band structure and density of states plots were generated using the open-source Python package, VaspVis \cite{Cal_12}.
\color{black}

\section{III. RESULTS and DISCUSSION}

\subsection{Film morphology and structural properties}
PtMnSb(001) surface exhibits a (2$\times$2) reconstruction, which is different from the (2$\times$1) structure reported for NiMnSb \cite{Barrier3}. Similarly to the III-V semiconductor cases \cite{IIIV_couting}, the surface reconstruction on a half Heusler (001) surface has been systematically discussed based on dimerization of the most electronegative element (typically $Z$ in $XYZ$) and vacancy formation for more electropositive elements following the electron counting rule \cite{HH_couting}. The longer Sb-Sb distance and the inclusion of more electronegative Pt compared to Ni may lead to different configuration and numbers of Sb-Sb dimers in PtMnSb. The surface of the PtMnSb films shows flat terraces with step height corresponding to the half unit cell of PtMnSb (Fig. 1(g)). 

Figure 2 summarizes structural characterization of PtMnSb films by XRD and STEM. The purpose of adopting LuSb barrier layer in growing PtMnSb on GaSb(001) is for preventing the interfacial reaction as well as providing a good template for PtMnSb nucleation. As shown in Fig. 2(a), direct deposition of PtMnSb on GaSb at 300$^{\circ}$C results in formation of interfacial Ga$_2$Pt as well as other decomposed impurity phases (marked with diamonds) such as Mn$_2$Sb and Pt. On the other hand, in the presence of a LuSb barrier, PtMnSb can be grown and annealed at elevated temperature for higher crystallinity without appearance of detectable additional phases. The cross-sectional STEM images in Fig. 2(b) reveals sharp interfaces between PtMnSb and LuSb as well as between LuSb and GaSb. We note that rare earth mono-pnictides have been reported to serve as a good diffusion barrier for other heterostructure systems between transitional metals and III-Vs \cite{Barrier1,Barrier2}. 

Reciprocal space maps measured around (115) Bragg peaks presented in Figs. 2(c)-2(e) show that PtMnSb films below 13 nm is grown pseudomorphically on LuSb-buffered GaSb, leading to a compressive strain up to 1.9\%. Accordingly, the out-of-plane lattice constant is elongated by 1.3\% following a Poisson ratio of 0.26. The rocking curve width of the strained PtMnSb is 0.13 deg., indicating much improved crystallinity compared to the previously reported PtMnSb films grown on oxides or metal-buffers \cite{PMS_film2,PMS_film3}. Further increasing the film thickness over 13 nm results in the lattice relaxation of PtMnSb. 

\begin{figure}
\begin{center}
\includegraphics[width=5.7cm]{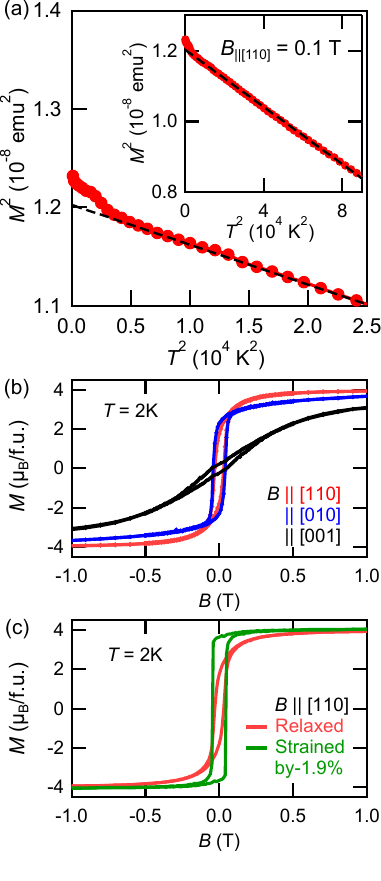}
\caption{
Magnetization properties of the PtMnSb films. (a) Temperature dependence of magnetization $M$ measured along the [110] direction shown by plotting $M^2$ vs. $T^2$. The inset presents a wider temperature range up to 300 K. The broken line shows the linear fit for the higher temperature region. (b) Magnetization curves of the 33 nm thick PtMnSb film taken at 2 K with field $B$ applied along the [110], [010], and [001] directions. (c) Magnetization curve of the compressively strained film (13 nm) compared to that of the relaxed films (33 nm).  
 }
\label{fig3}
\end{center}
\end{figure}

\subsection{Magnetization and magnetotransport}
Magnetic properties of half Heusler have been known to be affected by the underlying band structure and the chemical order \cite{Heusler_rev1}. While $T_{\mathrm{C}}$ of PtMnSb is well above the room temperature, we have observed anomalous increase of magnetization at the low temperature. Figure 3 (a) presents temperature dependence of magnetization ($M$) by plotting $M^2$ as a function of squared temperature $T^2$. $M^2$ varies linearly with respect to $T^2$ at higher temperatures (inset of Fig. 3(a)), while it deviates from the $T^2$ law and exhibits pronounced enhancement below 60 K. The $T^2$ law in the higher temperature region indicates a typical itinerant ferromagnetic behavior \cite{Spintheory}. On the other hand, anomalous enhancement of magnetization at low temperature could be attributed to suppression of spin flip (Stoner-type) excitation due to the development of a near-half-metallic state in the PtMnSb film as we discuss later. Similar deviation from the $T^2$ law at low temperatures has been observed for the bulk crystals of PtMnSb and half-metallic NiMnSb \cite{NiMnSb_transport,PMS_transport}. 

Magnetization curves measured at 2 K as function of field applied along the three different orientations are presented in Fig. 2(b). The 33 nm thick PtMnSb film shows the easy axis lying along the in-plane [110] direction with a small coercive field about 260 Oe. Such in-plane anisotropy is more reinforced by shape anisotropy in the thinner compressively-strained film where larger remanent moments and a coercive field about 400 Oe are observed (Fig. 3(c)). The saturation moment per formula unit (f.u.) reaches close to 4 $\mu_{\mathrm{B}}$ which is expected from the Slater-Pauling rule of half Heusler compounds, implying suppressed off-stoichiometry and chemical disorders in the PtMnSb films.

\begin{figure}
\begin{center}
\includegraphics[width=7cm]{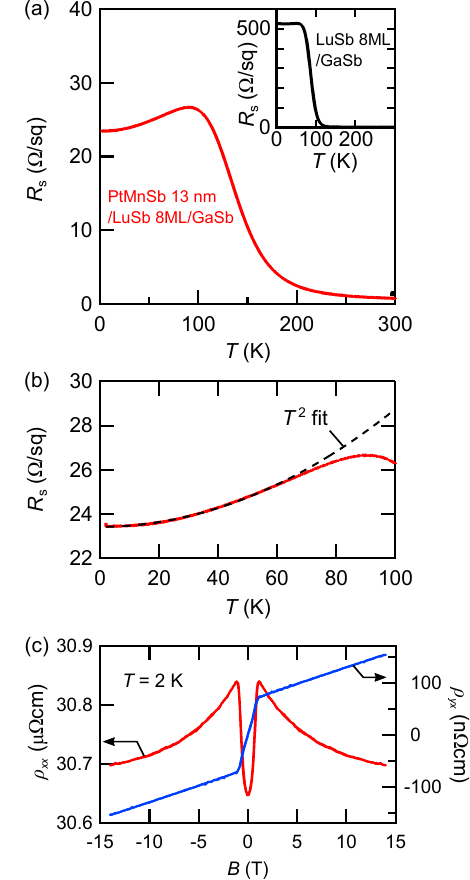}
\caption{
Transport properties of the PtMnSb films. (a) Temperature dependence of sheet resistance $R_{\mathrm{s}}$ for the PtMnSb 13nm/LuSb 8ML/GaSb sample. Inset shows the result for LuSb 8ML/GaSb sample. (b) Low temperature region of (a) obeying $T^2$ dependence (broken line). (c) Magnetoresistivity (left) and Hall resistivity (right) at 2 K with field applied along the out-of-plane direction.
}
\label{fig4}
\end{center}
\end{figure}

Figure 4(a) shows temperature dependence of the sheet resistance for a 13 nm thick PtMnSb film grown on 8 ML thick LuSb/GaSb. The conduction of LuSb buffered GaSb dominates over that of the PtMnSb film above 100 K and exhibits an insulating behavior. On the other hand, the low temperature metallic conduction appearing below the freezing out temperature of GaSb around 100 K can be associated with PtMnSb contribution. By evaluating the sheet resistance of a reference sample with only 8 ML LuSb grown on the GaSb substrate (inset of Fig. 4(a)), we have identified the conduction contribution of LuSb/GaSb is less than 5 \% of that of PtMnSb below 50 K. The sheetresistance of PtMnSb films at low temperature can be well-fitted by a $T^2$ law as shown in Fig. 4(b). This indicates that the spin disorder scattering is mainly caused by the spin wave excitation and not by the spin fluctuations involving spin reversal for which a $T^{5/3}$ law is expected \cite{Spintheory}. This is also in line with the picture of near-half-metallic state developing at the low temperature as suggested by the magnetization measurements in Fig. 3.

Figure 4(c) shows magnetic field dependence of longitudinal resistivity $\rho _{xx}$ and Hall resistivity $\rho _{yx}$ at 2 K. $\rho _{xx}$ exhibits little magnetoresistivity and $\rho _{yx}$ shows anomalous Hall effect accompanied by a tiny ferromagnetic hysteresis loop in the lower field region. The dominant carriers type is p-type, and its density is calculated to be 4.3$\times$10$^{22}$ cm$^{-3}$ by linear fitting of the ordinary Hall term above the saturation field. The heavily degenerate nature is consistent with the presence of large hole pockets around the Brillouin zone center as shown by the DFT calculations discussed in the following section.

\color{black}
\subsection{Electronic structure}

\begin{figure*}
\begin{center}
\includegraphics[width=1.0\textwidth]{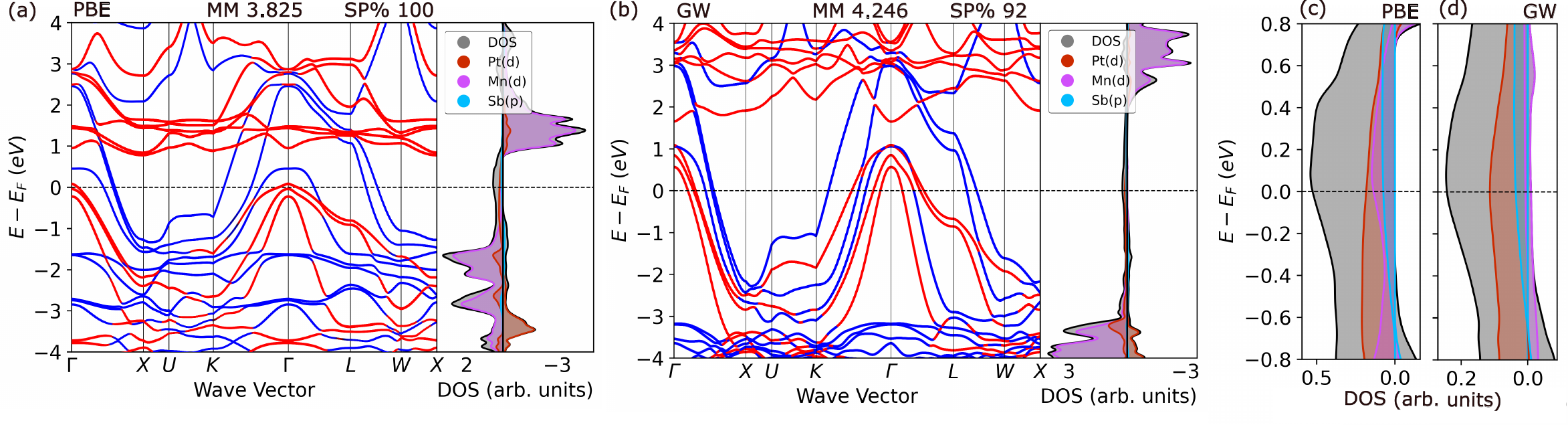}
\caption{
\color{black}Electronic structure of (strained) PtMnSb calculated using (a),(c) DFT with the PBE functional and (b),(d) QPGW. Spin-resolved band structures are shown with the majority bands colored in blue and minority bands colored in red. In the spin-resolved DOS, the majority channel is shown on the left and the minority channel on the right with contributions from selected states shown in different colors. Spin polarization (SP) and magnetic moment values (MM) are also shown. (c) and (d) present a magnified view of the DOS in the vicinity of the Fermi level for the PBE and the QPGW results, respectively.  
\color{black} }
\label{fig5}
\end{center}
\end{figure*}

Based on DFT calculations within the local density approximation (LDA), PtMnSb had been predicted to be a pure half-metal with the top of the spin minority bands situated right below the Fermi level \cite{PMS_SOC_cal, Cal_13}. The results of DFT calculations depend strongly on the choice of approximation for the exchange-correlation functional.  Local and semi-local functionals may fail to provide a reliable description of the magnetic properties of Heusler and half-Heusler compounds \cite{Cal_14, Cal_15, Cal_16,Cal_17,Cal_18,Cal_19,Cal_20,Cal_21,Cal_22,Cal_23,Cal_24}. Many-body perturbation theory within the $GW$ approximation, where $G$ represents the Green’s function and $W$ represents the screened Coulomb interaction, has been shown to provide a reliable description of Heusler and Half-Heusler compounds \cite{Cal_25,Cal_26, Cal_17}. In particular, the quasi-particle self-consistent GW (QPGW) scheme \cite{Cal_27, Cal_28, Cal_6} eliminates the dependence on the mean-field starting point of the non-self-consistent $G_0W_0$ scheme. It has been shown to perform well for magnetic and correlated materials \cite{Cal_17,Cal_29,Cal_30,Cal_31,Cal_32}.

Figure 5 shows a comparison of the spin-resolved band structure and density of states (DOS) of PtMnSb produced by DFT with the PBE functional and QPGW. The PBE functional produces a similar band structure to the LDA results reported in \cite{PMS_SOC_cal}, predicting PtMnSb to be a pure half-metal with 100\% majority spin polarization at the Fermi level and the minority bands lying just below the Fermi level at the $\Gamma$ point. The QPGW band structure is significantly different than the PBE band structure. The orbital decomposition of the DOS shows that the Mn $d$ orbitals shift downward in energy significantly upon switching from PBE to QPGW. This may be attributed to the self-interaction error (SIE) \cite{Cal_33} in (semi-) local DFT functionals, which manifests in shifting of highly localized states, such as the Mn $d$ orbitals higher in energy due to the spurious repulsion of the electron from its own charge density \cite{Cal_34,Cal_35,Cal_36,Cal_37}.  Around the Fermi level, the Pt $d$ states and the Sb $p$ states contribute significantly to the majority spin channel with both PBE and QPGW, as shown in Figs. 5(c) and 5(d). The difference between PBE and QPGW in the spin polarization around the Fermi level is mainly due to the Mn $d$ states, which contribute significantly to the majority channel with PBE, but with QPGW contribute to the minority channel instead. Although with QPGW PtMnSb is not a pure half-metal, it is still highly spin polarized with 92\% majority spin around the Fermi level. We refer to this as a near-half-metal.

\begin{figure}
\begin{center}
\includegraphics[width=8.7cm]{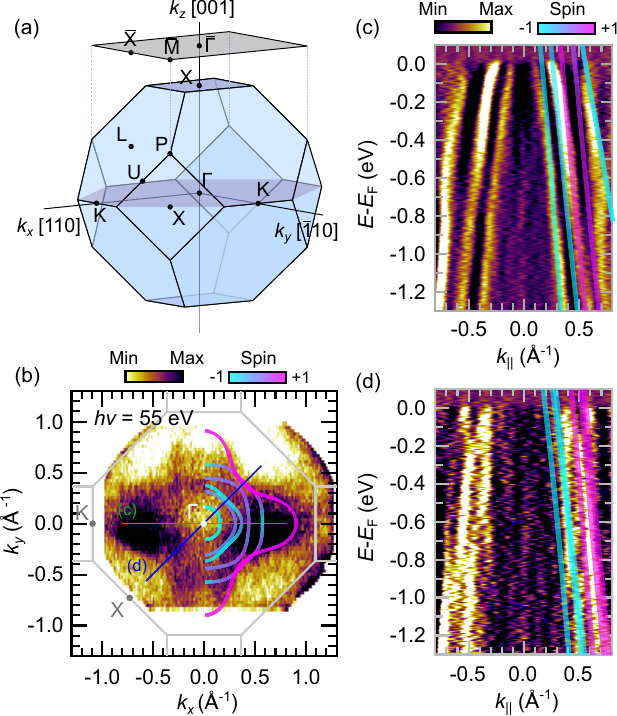}
\caption{\color{black}
ARPES measurements performed on the 5 nm thick PtMnSb(001) film. (a) Schematic of the bulk and the (001) surface Brillouin zone of PtMnSb. (b) Fermi surface cross-section with spin-resolved QPGW results superimposed on the ARPES data. Band dispersions along (c) $\bar{\Gamma}$-$\bar{X}$ and (d) $\bar{\Gamma}$-$\bar{M}$ directions measured near the bulk $\Gamma$ point using photon energy $h\nu = $ 55 eV. Curvature plots of the raw spectra are presented. QPGW results are superimposed on the ARPES data with the majority spin bands colored in magenta and the minority spin bands in cyan.\color{black}     
 }
\label{fig6}
\end{center}
\end{figure}

To experimentally clarify the electronic structure, we have performed ARPES measurements on a 5 nm thick PtMnSb film, which is compressively strained by 1.9 \%. We first present the Fermi surface map measured near the bulk $\Gamma$ point using incident photon energy of 55 eV. As shown in Fig. 6(b), the characteristic Fermi surface structure, where the outermost band exhibits a large hole pocket elongated along the fourfold bulk $\Gamma$-$K$ direction, is clearly captured. Figures 6(c) and 6(d) show the curvature plot of the band dispersions observed along the $\bar{\Gamma}$-$\bar{X}$ (parallel to bulk $\Gamma$-$K$) and $\bar{\Gamma}$-$\bar{M}$ (parallel to bulk $\Gamma$-$X$) directions (see Supplemental Material for the raw spectra \cite{SM}). The QPGW results are in good agreement with the ARPES data, though part of bands that are close to each other are not well-resolved in the experiment. A comparison of the PBE results to our ARPES data is also provided in the Supplemental Material \cite{SM}, showing discrepancies with experiment, in particular for the minority spin bands near the Fermi level. The QPGW results also agree better than the DFT-PBE results with the orbital-resolved PES data of Kang \textit{et al.} \cite{Cal_38}, as shown in the Supplemental Material \cite{SM}.  As the ARPES measurements are not spin-resolved, we rely on the QPGW calculations to assign spins to the states observed in ARPES. The computed band structure and Fermi level cross-section show six bands crossing the Fermi level along $\Gamma$-$X$ and along $\Gamma$-$K$. In the $\Gamma$-$X$ direction, the three minority bands are grouped together, followed by the three majority bands. In the $\Gamma$-$K$ direction, there are two minority bands, followed by two majority bands, followed by the remaining minority band and finally the third majority band. The latter is seen in the Fermi surface map in Fig. 6(b) but is outside the range of Fig. 6(c). Thus, we can conclude that PtMnSb is not pure half-metal but rather a near half-metal. Still, the spin polarization at the Fermi level of PtMnSb remains over 90 \%, which can effectively suppress the spin-flip excitation leading to the anomalous enhancement of magnetization at low temperatures similarly to the pure half-metal case.

\color{black}
\section{IV. CONCLUSION}
  
To conclude, we have demonstrated for the first time epitaxial stabilization of the half Heusler PtMnSb on a III-V substrate by molecular beam epitaxy. By adopting LuSb as a barrier layer, PtMnSb (001) films of high crystalline quality are grown on GaSb without interfacial diffusion or reactions. PtMnSb films exhibit a saturation moment of 4 $\mu_{\mathrm{B}}$/f.u. as expected from the Slater-Pauling rule of half Heusler compounds, and an anomalous increase of magnetization at low temperature similar to the half-metallic NiMnSb. \color{black}Previous DFT-LDA calculations \cite{PMS_SOC_cal} and DFT-PBE calculations performed here predict a half-metallic gap located close to the Fermi level. In contrast, the QPGW scheme within many-body perturbation theory   shows  that PtMnSb is rather a near-half-metal with the bands of both spins crossing the Fermi level, but with a majority spin polarization of over 90 \%. The QPGW band structure is in good agreement with ARPES measurements. Despite not being a pure half-metal, the high spin polarization in PtMnSb in combination with its high SOC and inversion-symmetry-broken structure, could provide a useful platform for spintronics. In this respect, our demonstration of highly crystalline PtMnSb films grown on a III-V substrate provides opportunities for further investigations of its magnetic properties, such as spin orbit torques \cite{PMS_strain2} and other spin-dependent transport. \color{black}

\color{black}
 This work was supported by the U.S. Department of Energy (contract no. DE-SC0014388). Work at CMU was supported by the Department of Energy through grant DE-SC-0019274.  We acknowledge the use of shared facilities of the NSF Materials Research Science and Engineering Center (MRSEC) at the University of California Santa Barbara (DMR-2308708). Use of the Stanford Synchrotron Radiation Lightsource, SLAC National Accelerator Laboratory, was supported by the U.S. DOE, Office of Science, Office of Basic Energy Sciences under Contract No. DE-AC02-76SF00515. We also acknowledge the support of the U.S. Department of Energy, Office of Science, Office of Basic Energy Sciences, Division of Material Sciences and Engineering, under Contract No. DE-AC02-76SF00515. This research used computing resources of the National Energy Research Scientific Computing Center (NERSC), a U.S. Department of Energy Office of Science User Facility operated under Contract No. DE-AC02-05CH11231. We would like to thank S. Khalid, A. Janotti, S. Chatterjee, W. K. Peria, and P. A. Crowell for fruitful discussions.
\color{black}

%
\clearpage
\end{document}